\documentclass[twocolumn,aps,superscriptaddress]{revtex4}
\usepackage{graphicx}

\begin{document}

\title{A multiscale study of electronic structure and quantum transport in $C_{6n^2}H_{6n}$-based graphene quantum dots}

\author{I. Deretzis}
\email[]{ioannis.deretzis@imm.cnr.it}
\affiliation{Scuola Superiore, Universit\`{a} di Catania, I-95123 Catania, Italy}
\affiliation{CNR-IMM, I-95121 Catania, Italy}

\author{G. Forte}
\affiliation{Dipartimento di Scienze Chimiche, Universit\`{a} di Catania, I-95126 Catania, Italy}

\author{A. Grassi}
\affiliation{Dipartimento di Scienze Chimiche, Universit\`{a} di Catania, I-95126 Catania, Italy}

\author{A. La Magna}
\affiliation{CNR-IMM, I-95121 Catania, Italy}

\author{G. Piccitto}
\affiliation{Dipartimento di Fisica e Astronomia, Universit\`{a} di Catania, I-95123 Catania, Italy}

\author{R. Pucci}
\affiliation{Dipartimento di Fisica e Astronomia, Universit\`{a} di Catania, I-95123 Catania, Italy}

\date{\today}

\begin{abstract}
We implement a bottom-up multiscale approach for the modeling of defect localization in $C_{6n^2}H_{6n}$ islands, i.e. graphene quantum dots with a hexagonal symmetry, by means of density functional and semiempirical approaches. Using the \textit{ab initio} calculations as a reference, we recognize the theoretical framework under which semiempirical methods describe adequately the electronic structure of the studied systems and thereon proceed to the calculation of quantum transport within the non-equilibrium Green's function formalism. The computational data reveal an impurity-like behavior of vacancies in these clusters and evidence the role of parameterization even within the same semiempirical context. In terms of conduction, failure to capture the proper chemical aspects in the presence of generic local alterations of the ideal atomic structure results in an improper description of the transport features. As an example, we show wavefunction localization phenomena induced by the presence of vacancies and discuss the importance of their modeling for the conduction characteristics of the studied structures.
\end{abstract}

\maketitle

\section{Introduction}
Graphene is a carbon allotrope material that has triggered a vast interest within both academic and industrial communities for its peculiar electrical, mechanical and optical characteristics\cite{2007NatMa...6..183G}. Additionally, graphene results  extremely convenient in terms of electronic structure modeling, since its planar monolayer topology and $sp^2$ hybridization allow for a simple and accurate $\pi$-orbital tight-binding (TB) description of its electronic bands\cite{PhysRev.71.622,PhysRevB.66.035412}. This approach has been extensively implemented also in the case of quantum transport studies of quasi one-dimensional graphene constrictions with armchair or zigzag edges, widely known as graphene nanoribbons (GNRs)\cite{magna:153405,2007IEDL...28..760F,2008arXiv0805.0035L,neophy08}. In the course of these few years since graphene's laboratory isolation and due to the intense research on the field, experimental knowledge has grown; e.g. recent transmission electron microscopy images of monoatomic graphene flakes obtained with mechanical exfoliation have shown Stone-Wales defects and vacancies on the crystal membrane\cite{Meye08}, while chemical reactivity with metal oxides has been achieved both in defected and in edge sites\cite{Wang08}. Theory on the other hand by means of first principles calculations has demonstrated energetically favorable chemisorption processes on the areas that diverge from the perfect two-dimensional atomic lattice\cite{Bouk08}. The aforementioned advances, among others, have raised new challenges in the field of graphene-based modeling in chemically and geometrically complex environments. Ideally, density functional theory (DFT) with appropriate exchange-correlation functionals could formulate an accurate and transferable framework for simulations on these systems \cite{son:216803,son:216803,2008PhRvB..77o3411H}. The undoubted intrinsic rigor of \textit{ab initio} approaches comes in hand though with computational overload limitations. In this sense, only order $\sim 10^3$ can be affordable for all-electron bandstructure calculations (under linear scaling optimization techniques\cite{MatthewC,izmaylov:104103,1996CPL...253..268W,Scuseria1999}), while an extra load has to be considered for self-consistent nonequilibrium quantum transport studies. Therefore, it becomes evident that a massive atomic reconstruction study in the presence of generic local alterations of the symmetry or the ideal atomic structure cannot be addressed with `dogmatic' singular theoretical approaches (e.g. pure \textit{ab initio} methods, pure TB studies etc.), whereas a versatile multilevel approach would seem more appropriate.

\begin{figure}%
\centering
\includegraphics[width=0.7\linewidth]{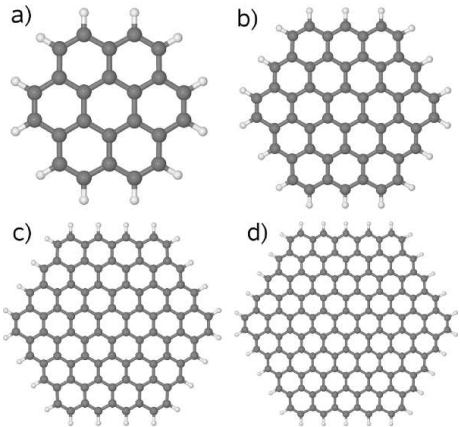}
 \caption{$C_{6n^2}H_{6n}$ molecular complexes: a) $C_{24} H_{12}$ (coronene, n=2), b) $C_{54} H_{18}$ (coronene 19, n=3), c) $C_{96} H_{24}$ (coronene 37, n=4), d)  $C_{150} H_{30}$ (coronene 61, n=5)}
 \label{fig:coronenes}
\end{figure}

In this article a multiscale study of electronic structure and quantum transport is carried out for graphene quantum dot complexes. The study's groundwork focuses on: (i) the structural characteristics, and (ii) the methodological approach. The islands under consideration are the coronene molecules\cite{1966RSPSA.289..366F,2008PhLA..372.6168F} with a general chemical type of $C_{6n^2}H_{6n}$ in a pure, defected (with a single vacancy) and hydrogen functionalized form. These can be thought of as planar complexes of benzene rings that grow rotating around a central benzene ring, forming six hydrogen-passivated zigzag edges (see fig. \ref{fig:coronenes}). Methodologically, electronic structure is initially studied with DFT while optical properties are calculated within the time-dependent density functional theory (TD-DFT). These results serve as a reference for the comparison with similar calculations by means of two parameterized semiempirical methods: (i) the extended H\"{u}ckel (EH) theory \cite{hoffmann:1397} and (ii), the next-neighbor tight-binding model. As soon as a proper functional framework is identified for the semiempirical approaches the study proceeds with the calculation of electronic transport. Particular attention is paid to the effect of defect localization both within a level of characterization as well as model calibration, since the computational results indicate impurity-like behavior of single vacancies in these systems. Conceptually, although the final objective is physical (quantum transport modeling in graphene islands), the basis is founded on chemistry (multiscale comparative analysis of the electronic structure).

The paper is organized as follows: In section \ref{section2} we review the methodological steps of the theoretical model,  section \ref{section3} presents first principles and semiempirical electronic structure calculations for $C_{6n^2}H_{6n}$ islands up to $n=5$, section \ref{section4} analyzes transport characteristics and focuses on the effects of local atomic reconstruction on the conductance, while in section \ref{section5} we discuss our results.

\section{Methodology}
\label{section2}
Geometry relaxation and electronic structure properties (eigenvalues, eigenfunctions, density of states) of various $C_{6n^2}H_{6n}$ clusters are extrapolated by DFT calculations on a split-valence double-zeta (3-21g \cite{3-21g-1,3-21g-2,3-21g-3}) and a minimal (STO-3G \cite{hehre:2657,collins:5142}) basis set, as implemented in the GAUSSIAN code\cite{g03}. The semiempirical three-parameter hybrid nonlocal exchange and correlation functional of Becke and Lee, Yang and Parr\cite{PhysRevB.37.785,becke:5648,1980CaJPh..58.1200V,Steph94} (B3LYP) has been chosen here for its capacity to predict a large range of molecular properties for aromatic systems\cite{Mode06,Bouz05}. Additional optical properties (excitation energies, fundamental optical gaps) are calculated within a TD-DFT approach for comparison between theory and experiment. Electronic structure results are then confronted with similar ones obtained by two semiempirical methods\cite{124,Dere06} that also present a precision/efficiency mismatch among them: (i) the extended H\"{u}ckel method, and (ii) a next-neighbor tight-binding model. In the case of the EH method three distinct parameterizations are used: (i) the first one considers a standard valence $2s2p$-basis set of single-$\zeta$ Slater orbitals for C atoms, principally deriving from the initial values used by Hoffmann\cite{hoffmann:1397}(EH-sp from now on)\footnote{$Eon-site(C_s)=-21.4eV$, $Eon-site(C_p)=-11.4eV$, $Eon-site(H)=-13.6eV$, $\zeta(C_s)=1.625$, $\zeta(C_p)=1.625$, $\zeta(H)=1.3$, $coeff(C_s)=1$, $coeff(C_p)=1$, $coeff(H)=1$, $K=1.75$}. (ii) The second one is a $2s2p3d$-based parameterization with valence/polarization double-$\zeta$ exponents and C parameters fitted to recreate the bandstructure of two-dimensional graphene as given by DFT calculations\cite{Cerdaweb,kienle:043714,PhysRevB.61.7965} (EH2-spd from now on). (iii) The third parameterization derives in the similar way to the second one, whereas here the polarization orbitals are absent\cite{Cerdaweb,kienle:043714}(EH2-sp from now on). For the first-neighbor TB model a standard $t_{0}=2.7eV$ C-C hopping integral is used, while vacancies are approximated with the insertion of a local point potential $U\rightarrow \infty$ (unless explicitly referred to in the text). Although all methods construct the molecular orbitals on the basis of the linear combination of atomic orbitals there is a distinct difference in the level of accuracy that each method delivers. In the DFT case the basis set is comprised of Gaussian-type orbitals with weighting coefficients that are both calculated self-consistently in order to reproduce the best approximation of the exact ground state density of the system. In the EH case the bases are nonorthogonal Slater-type orbitals with fixed weighting coefficients that have been parameterized on the basis of experimental data or first principles calculations. Finally in the TB case no real orbitals exist and the system Hamiltonian is constructed by a simple finite difference approach on next-neighbor atoms that through a proper choice of the hopping integral is representative of the $\pi$-orbital in the $sp^2$ hybridization scheme. Naturally, the level of computational efficiency is the inverse, ranging from molecular (DFT) to mesoscopic (TB). 

Electronic structure results are obtained through a direct diagonalization of the respective Hamiltonian matrix. Comparisons take place in terms of highest occupied molecular orbital (HOMO) and  lowest unoccupied molecular orbital (LUMO) gaps, energy eigenstates $\epsilon_\alpha$ and their respective eigenfunctions $\Psi_\alpha$. The local density of states LDOS($\vec{r},E)$ at the positions $\vec{r}$ of the device atoms at energy $E$ is calculated as:
\begin{equation}
 LDOS(\vec{r},E) = \sum_\alpha{ | \Psi_\alpha (\vec{r}) |^{2} \delta(E-\epsilon_\alpha) },
\end{equation}
where $\delta$ is the Delta function, while summing over all atoms gives the total density of states (DOS) of the molecular systems at this energy.

Quantum transport is calculated within the non-equilibrium Green's function formalism\cite{2000SuMi...28..253D}. In particular, the method is based on the single particle retarded Green's function matrix $ G = [ES - H - \Sigma_L - \Sigma_R]^{-1}, $
where $E$ is the energy, $H$ and $S$ are the device Hamiltonian and the overlap matrix respectively (written in an appropriate basis set), while $\Sigma_{L,R}$ are the self-energy matrices that account for the effect of scattering due to the left $(L)$ and right $(R)$ contacts. In the TB case the overlap matrix coincides with the unitary one. The $\Sigma_{L,R}$ terms can be expressed as $\Sigma = {\tau}g_s{\tau}^\dagger$, where $g_s$ is the surface Green function specific to the contact type and $\tau$ is the Hamiltonian relative to the interaction between the device and the contact. The calculation of the Green's function permits for the evaluation of all the quantities of interest for conduction, e.g. the device spectral function is the anti-hermitian part of the Green's function $A = \imath(G-G^\dagger)$ from which the Density of States can be obtained as $D(E) = \frac{1}{2\pi} Trace(AS)$. Moreover in the coherent transport regime, the expression used for the zero-bias transmission probability reads $T(E)=Trace({\Gamma_L}G{\Gamma_R}G^\dagger)$, where $\Gamma_{L,R}=\imath (\Sigma_{L,R}-\Sigma_{L,R}^\dagger)$ are the contact spectral functions.

In this study $C_{24} H_{12}$, $C_{54} H_{18}$, $C_{96} H_{24}$ and $C_{150} H_{30}$ complexes have been considered in their pure, defected (with a single vacancy) and hydrogen functionalized form. All structures have been relaxed by DFT molecular dynamics while relaxation information is also used by the EH method. In the case of TB an ideal reconstruction of the molecular structure is considered since the latter does not account for interatomic distances. Finally, for the quantum transport calculations the islands are placed within two semi-infinite $Au(111)$ metallic planes (directly considered in the case of EH, appropriately fitted in the case of TB\cite{Dere06}) in a molecular bridge configuration.

\section{Comparative analysis of the electronic structure between first-principles and semiempirical methods}
\label{section3}
A proper treatment of quantum transport modeling has to take care of both quantitative and qualitative aspects of the electronic structure of a molecular system. In this sense, if the value of the HOMO-LUMO gap is a quantitative feature, the form of the HOMO and LUMO wavefunctions, or similarly, the local density of states of the structure for energies near the HOMO/LUMO states are qualitative characteristics. It can be argued that in terms of conduction, although the former can influence scaling, the latter can affect the shape of the current-voltage curve. In addition, higher-bias conduction requests accuracy for entire conduction/valence bands. Such considerations imply that a proper description of the local density of states of a molecular system by means of a quantum chemical method can be fundamental for the correct modeling in terms of quantum transport. This section examines electronic configuration aspects one by one.

\subsection{Geometry relaxation}
From a numerical point of view, distance between the atomic sites influences the electronic structure of a molecular system by affecting both overlap and Hamiltonian matrix elements. Geometry relaxation with the DFT method has shown that near the island edges, complex distance polymerization effects can be observed that tend to periodically increase/decrease C-C bonding for about $5$\%{} from the equilibrium distance ($C-C_{equil}\approx 1.42$\AA). Such feature tends to propagate also in the inner parts of the clusters albeit in a continuously decreasing extent, while only the central benzene rings result having equal interatomic C distances. Distance shortenings due to hydrogen passivation have been also observed in the case of graphene nanoribbons\cite{son:216803}, whereas the effect there is localized near the edges. From a methodological point of view, variable bond-lengths can have a practical consequence in the parameterization of the hopping integral for the TB method, where for accuracy's sake an evaluation of each atomic pair distance should take place prior to the assignment of the integral value (in this sense the method becomes similar to the single $\pi$-orbital H\"{u}ckel one). If such information is not available inaccuracies in the TB Hamiltonian can occur. It can be argued that futher complications in the correct estimation of interatomic distances have to be considered in the case of interaction between the molecular structures and a substrate (e.g. for the reproduction of laboratory conditions\cite{2002JAP....91.9095S,1998JAP....84..268C,Zimmermann1992296}). In this case combined melecular/substrate atomistic modelling should enlight interface bonding interactions for both the chemical and the structural characteristics.

\subsection{Energy levels}

\begin{figure}%
\centering
\includegraphics[width=0.6\linewidth, angle=-90]{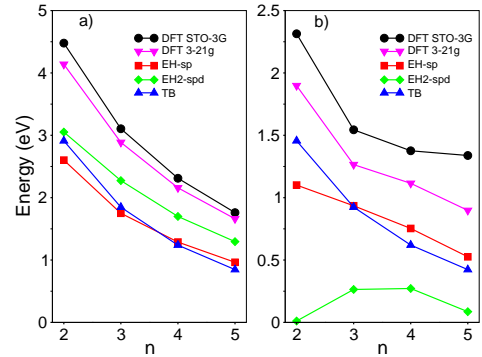}
 \caption{HOMO-LUMO gaps of pure (a) and defected (b) $C_{6n^2}H_{6n}$ quantum dots by means of DFT, EH and TB.}
 \label{fig:gaps}
\end{figure}

A well-known aspect of geometrical symmetry is the presence of orbital degeneracies. Such feature is captured by all methods for the pure structures, from DFT to TB. Eventually, symmetry breaking events (e.g. the presence of a single vacancy) lift such degeneracies and split the respective energy levels. In this subsection we visualize characteristics of the eigenvalue spectrum in a full quantum scale, in terms of energy gaps and state alignment over the energy axis. Figure \ref{fig:gaps} shows HOMO-LUMO gaps for pure and defected structures by means of DFT (both 3-21g and STO-3G), EH2-spd, EH-sp and TB, while EH2-sp results are similar to EH2-spd and are not shown. Regarding the pure structures the following observations can be made: (i) the gap value given by the DFT is bigger than those obtained by the semiempirical methods. This aspect is not directly related to the exchange-correlation absence in the semiempirical cases (since parameterization can take place on the basis of \textit{ab initio} calculations) but with the chemical environment considered for the parameterization, which usually considers bulk structures where the effect of confinement cannot be evaluated. It is moreover interesting to evidence that the minimal basis set in the DFT method slightly overestimates this value with respect to the 3-21g case. (ii) The EH2-spd method with its orbital foundation and $sp^2$-hybridized calibration approximates better the DFT results with respect to the other semiempirical methods. (iii) Albeit the clear difference in terms of the methodology, results by EH-sp and TB are very similar. It should be noted here that a rough method to bring semiempirical models closer to first principles calculations is by globally fitting the hopping integral (or similarly the Wolfsberg-Helmholtz $K$ constant for EH). However in this case the parameterization looses any meaning outside the designated geometrical environment.

The validity of the results obtained by the first-principles B3LYP/32-1g model for these structures with respect to the semiempirical methods is tested by direct comparison with experimental data on the fundamental optical gap of the $C_{24} H_{12}$ island. For this purpose we follow a TD-DFT approach for the calculation of the excitation energies of this cluster. The calculated value for the fundamental optical gap is $E_{opt}=3.18eV$, which is in a good agreement with the experimental value of $E_{opt}=3.29eV$ measured in Ref. \cite{2002JAP....91.9095S}\footnote{We have confirmed the good TD-DFT estimation of the optical gap with respect to the experimental value also by calculating excited state energies within the Configuration Interaction Singles (CIS) Method. The optical gap given by the latter is $E_{opt}=4.19eV$, which is by far bigger than the experimental and the TD-DFT value.}. The difference between the ground state ($E_{HOMO-LUMO}=4.13eV$) and the excited state ($E_{opt}=3.18eV$) is also consistent with experimental measurements for similar structures, since the exciton binding energy in these complexes gives rise to a $0.5-1 eV$ reduction of the optical gap with respect to the ground state HOMO-LUMO gap\cite{Hill2000181}.

Moving on to the defected structures (with a single vacancy in the central benzene ring) an expected reduction of the gap value can be observed, while DFT basis set differences become more pronounced. For the semiempirical methods the picture changes qualitatively only in the EH2-spd case. Considering spin-degeneracy, the EH2-spd parameterization assigns the HOMO and the LUMO states to two quasi-degenerate levels prior to the real gap, which in this case is represented by the LUMO and LUMO+1. A detailed study of the corresponding eigenvectors (with respect to eigenvectors given by the DFT and EH-sp) shows that in the case of the defected structures an incorrect state is inserted at the energy axis inside the energy gap. In this sense, although the EH2-spd/EH2-sp parameterizations demonstrate overall optimal characteristics (e.g. see ref. \cite{kienle:043714} for a study on carbon nanotube band structure and the next subsection for pure clusters), a careful use might be necessary for transport calculations in defected coronene systems.

\begin{figure}%
\centering
\includegraphics[width=0.7\linewidth, angle=-90]{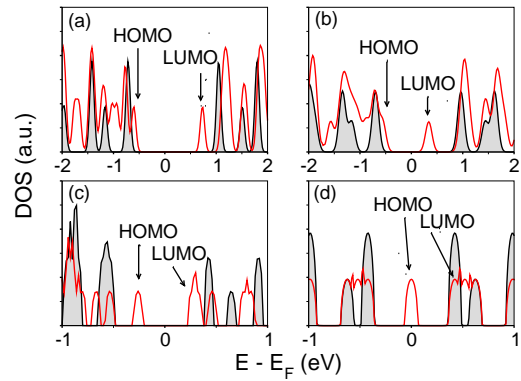}
 \caption{Density of states around the Fermi level for pure (black line) and defected (red line) $n=5$ islands by means of (a) DFT (STO-3G), (b) DFT (3-21g) (c) EH-sp and (d) TB. In all cases, a small smearing has been applied.}
 \label{fig:dos}
\end{figure}

Another important aspect that regards energy eigenstates is their alignment over the energy axis, moreover when local alterations of the symmetry `break' the ideal atomic structure. For few-atom molecular complexes a neat way to visualize this is with their density of states spectrum as a function of energy. Figure \ref{fig:dos} plots DOS functions for pure and defected $n=5$ islands by means of DFT, EH-sp and TB. In this case the lifting of the symmetry-induced degeneracy inserts states that tend to `shrink' the band gap. In the case of TB, a state always appears in the center of the pure structure's gap. This well-known effect has its origin at the bipartite nature of the honeycomb lattice, where the presence of a single vacancy in one of the two sub-lattices inserts a zero energy mode at the Fermi level of the system (i.e. at energy E=0)\cite{pereira:115109}. The key issue though arising  from figure \ref{fig:dos} is that for the more sophisticated methods the HOMO state is not located in the center of the pure structure's gap, but shifted towards lower energies next to the valence band. Such feature is captured by both DFT and EH-sp, although in a quantitative disagreement. The analysis therefore implies that TB gives a rigid picture of the gap state with respect to more sophisticated models. Under this perspective, this study will try to affront the problem by introducing a further parameterization for the point defect (see section \ref{section4}).

\subsection{Qualitative evaluation: molecular orbitals}

A most important aspect for transport in nanostructures is the availability of states (either full or empty) within the conduction window. In conjunction, a very important factor for the correct treatment of conduction are the eigenvectors that correspond to these states, from which topological features can be deduced (e.g. localization, polarization etc.). In this context the semiempirical methods have been evaluated on the basis of DFT results for pure, defected and hydrogen functionalized structures. Results shown here are for the DFT 3-21g, EH-sp, EH2-sp and TB models. A qualitative correspondence has been obtained for the DFT STO-3G and the EH2-spd bases with respect to their method counterparts\footnote{The surprisingly accurate correspondence in terms of molecular orbitals obtained for the 3-21g and STO-3G bases within the DFT scheme indicates that the minimal basis set can be used in disordered graphene-based systems with only small quantitative compromises (see paragraph \ref{section3}.B).}.

\begin{figure}%
\centering
\includegraphics[width=0.7\linewidth]{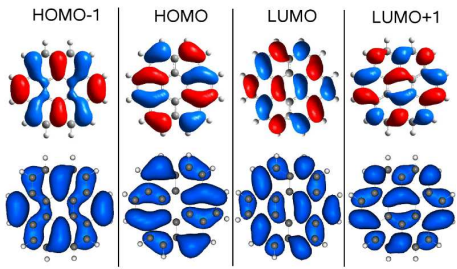}
 \caption{Molecular orbitals for the HOMO-1, HOMO, LUMO and LUMO+1 states of the $n=2$ complex by means of DFT (upper) and  EH-sp (lower).}
 \label{fig:coroDFT-EH}
\end{figure}

\subsubsection{Pure structures}
TB: In the absence of real atomic orbitals and with the restriction of its limited basis set the TB method demonstrates a progressive wavefunction descriptive capacity from smaller to larger complexes. In particular TB eigenvectors for conduction/valence eigenstates have a poor resemblance with the respective DFT ones for the $C_{24} H_{12}$ molecule, while similarity becomes gradually better for bigger structures. This behavior is due to the C-C bonding distance polymerization features discussed earlier, which have a higher impact for the smaller complexes. Arriving at the $n=5$ complex, DFT-TB matching becomes adequate, hence, the TB method is qualified for the electronic structure description of these structures with a respective number of C atoms and onwards.

EH-sp: The method is in a qualitative agreement with the DFT one only for the energy degenerate HOMO/HOMO-1 and LUMO/LUMO+1 pairs for all studied structures (see fig. \ref{fig:coroDFT-EH} for the coronene molecule). Moving away from these states towards the valence band accuracy is lost, not in the form of the wavefunctions whereas in the correct order that these appear. Conduction band description results poor.

\begin{figure}%
\centering
\includegraphics[width=0.7\linewidth]{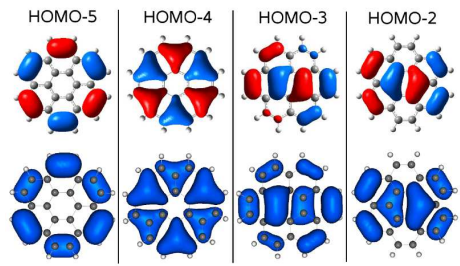}
 \caption{Molecular orbitals for the HOMO-5, HOMO-4, HOMO-3 and HOMO-2 states of the $n=2$ complex by means of DFT (upper) and EH2-sp (lower).}
 \label{fig:coro2DFT-EH}
\end{figure}

EH2-sp: Matching between EH2-sp and DFT wavefunctions is excellent for all pure islands and for both valence and conduction energy zones (e.g. see figure \ref{fig:coro2DFT-EH} for the valence band of $C_{24}H_{12}$). It is evident here that the chemical environment in which the parameterization has taken place (bulk graphene) and the double-exponent Slater orbitals play a crucial role in the representation of correct molecular orbitals. It is also interesting to note the mismatch in the results obtained by the EH2-sp and the EH-sp parameterizations, even if the quantum chemical method is the same.

Finally, magnetism issues that could arise due to the presence of zigzag terminated edges in these complexes are not confirmed by the the \textit{ab initio} calculations, contrary to zigzag GNRs\cite{son:216803,2008PhRvB..77o3411H,1996JPSJ...65.1920F} or coronene islands above a critical size (based on mean-field Hubbard model calculations  \cite{2007PhRvL..99q7204F}). It can be therefore stated that the absence of a self-consistent exchange evaluation in the semiempirical methods does not compromise the obtained results with respect to the DFT case in the present study.

\subsubsection{Defected structures}
\begin{figure}%
\centering
\includegraphics[width=0.8\linewidth]{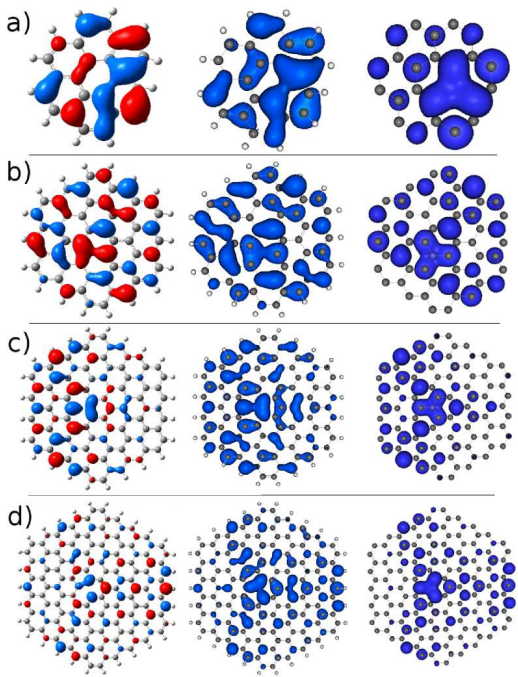}
 \caption{Highest occupied molecular orbitals for a) n=2, b) n=3, c) n=4 and d) n=5 islands with a single vacancy by means of DFT (left), EH-sp (middle) and TB (right). The TB orbital representation is purely demonstrative by assigning Slater type $p_z$ orbitals with the same parameters as the EH method.}
 \label{fig:HOMOvac}
\end{figure}

The presence of a single vacancy in these islands provokes a distortion of the atomic structure since the remaining $\sigma$ dangling bonds tend to recombine by leaving their equilibrium positions. On the other hand, $\sigma$-orbital energies are too far away from the HOMO-LUMO states and do not contribute to the formulation of the respective wavefunctions. Most importantly, apart from a symmetry breaking effect in topological terms, the presence of the vacancy imposes a localization of the wavefunctions that correspond to the various eigenstates, making such complexes `sensitive' to the positioning of a nanoprobe. In terms of the various methodologies we have obtained:

TB: The lack of information concerning distance in the TB method is even more important for the defected structures, where the smaller the structure the higher is the effect of the vacancy on its deformation. In this sense the TB method with its standard parameterization is inadequate for the description of the electronic structure of these complexes, whereas like in the case of pure structures, description gradually betters as the complexes grow, with the following particularities: (i) HOMO wavevectors present a succession of zero and non-zero values for neighboring atomic sites and (ii) for even $n$-indexed molecules ($n=3$, $n=5$ etc.) the hexagonal edge that corresponds to the defected site presents atoms with zero LDOS for the HOMO eigenstate (fig. \ref{fig:HOMOvac}). This last observation is crucial in terms of transport modeling and its implications will be discussed in the next section.

EH-sp: It describes better the valence (HOMO, HOMO-1, HOMO-2) than the conduction band. Moreover, moving towards the bigger structures accuracy is increased and for the $n=5$ island description becomes adequate for the valence and discrete for the conduction band.

EH2-sp: The main drawback of this parameterization has to do with the presence of an incorrect HOMO eigenstate as discussed in the previous subsection. Overall it offers a valid alternative to the DFT results, on the other hand though, the importance of the HOMO state in terms of conduction modeling requires attention in its use in defected graphene environments. 

Finally a remark on the $C_3$ point symmetry of the HOMO wavefunction around the vacancy should be made\cite{pereira:115109} (feature that is captured by all methods), where clearly a non-zero magnetic moment arises (also obtained with Hartree-Fock-based calculations on the same complexes\cite{2008PhLA..372.6168F}). 

\subsubsection{Hydrogen-functionalized structures}
Results for the complexes where the defected site has been functionalized by a hydrogen atom that saturates one $\sigma$ dangling bond do not differ substantially from their nonfunctionalized counterparts. Here the role of hydrogen slightly influences the structure's geometrical relaxation whereas wavefunctions are similar to non-passivated molecules with defects, as $\sigma$-orbital energies are too far away from the zone of interest for conduction. Consequently the discussion made in the previous subsection is valid also in this case for the methods that directly account for the presence of hydrogen (DFT, EH).

\subsection{Conclusion}
Semiempirical models in graphene-based quantum dot structures can be successfully used within a certain framework that is established by their quantum chemical limitations. The Extended H\"{u}ckel method with its real-orbital foundation can cope with a great number of qualitative features, whereas the role of parameterization proves to be fundamental. In this sense EH2-spd/EH2-sp are excellent alternatives to DFT for pure coronene structures, whereas defected/functionalized complexes are more appropriately treated by the EH-sp model. On the other hand, TB with the standard parameterization can be used for the study of conduction in large defect-free systems, while modeling remains a challenge for defected complexes since results appear too `radical'. In this sense a further parameterization of the defected site within its particular topological environment is necessary for the correct estimation of the electronic structure, argument that will be treated in the next section.

\section{Quantum transport}
\label{section4}

\begin{figure}%
\centering
\includegraphics[width=0.7\linewidth]{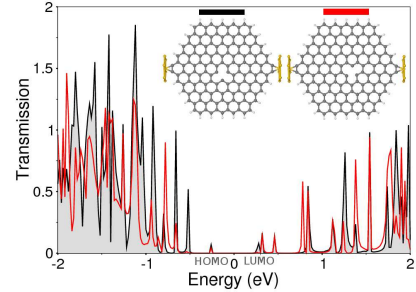}
 \caption{Transmission as a function of energy by means of EH-sp for the defected $n=5$ complex, for two equivalent contact configurations that differ only in the position with respect to the defected site.}
 \label{fig:cp3Trans}
\end{figure}

The study's bottom line is to efficiently model transport phenomena on graphene-based quantum dot systems respecting the chemical aspects that arise due to the particularity of the chemical/geometrical environment. The importance of a proper description of the electronic structure on conduction can be better appreciated in the case of defected islands, where the presence of the vacancy is a reason for topological asymmetries also on the formation of the molecular orbitals (see fig. \ref{fig:HOMOvac}). In this section, a numerical analysis takes place for the defected $n=5$ complex initially with the EH-sp method, while results are used for a critical evaluation of similar calculations made with the TB model. Two equivalent molecular bridge configurations are used, where the source-device-drain geometry differs only in the position of the contacts with respect to the vacancy site (fig. \ref{fig:cp3Trans}). In detail, two opposite edge corners of the aforementioned dot have been inserted between two semi-infinite Au(111) metallic planes, which model the metallic probes of an atomic force microscope. The contact Hamiltonian is also written within the EH theory using an appropriate $spd$ basis\cite{Dere06}. A prerequisite of equivalence for the contact bonding between the two configurations is explicitly requested for an evaluation of transport without geometrical or bond strength implications\cite{Dere06}. Here, contacts are $1.7$\AA{} distant from the edge C atoms (avoiding strong invasiveness) and are geometrically symmetrical with respect to the molecular structure. The transmission probabilities obtained for the two configurations have a distinct character, whereas differences are not fundamental for the conduction characterization of the system. Namely, both configurations give a non-zero transmission value corresponding to the HOMO state, whereas differences exist both in the valence and conduction band. The divergences can only be attributed to the different geometrical positions of the contacts with respect to vacancies that reflect unequal interface chemical bondings due to orbital localization phenomena. The minor impact of such phenomena on the conduction characteristics is driven by the real atomic orbital foundation of both contact and device wavefunctions that constitute bonding interactions that exceed next neighbor distances. Therefore, e.g. if local disorder provokes a nullification of the LDOS at the contact-device interface at a certain energy, transmission is still possible if this zero LDOS expands in a smaller area than that of orbital overlap between contact and further device atoms with a finite LDOS. The same concept can be described from a quantum mechanical perspective, where the presence of the contacts induces a constant perturbation on the bare device's Hamiltonian and the effective Hamiltonian now writes:
\begin{equation}
 \hat{H}_{eff}= \hat{H}_0 + \hat{H}_{0L} + \hat{H}_{0R}
\end{equation} 
Here $\hat{H}_0$ corresponds to the molecular Hamiltonian in the absence the contacts and $\hat{H}_{0L,0R}$ are the Hamiltonian components that arise due to the interaction between the device and the left/right contact. According to EH theory for the localized HOMO state we get:
\begin{equation}
 < \Psi_{HOMO} | \hat{H}_{0L} | \Psi_{HOMO} > \neq 0
\end{equation} 
and
\begin{equation}
 < \Psi_{HOMO} | \hat{H}_{0R} | \Psi_{HOMO} > \neq 0
\end{equation} 
This finite value of both integrals makes transport plausible from the HOMO state.

\begin{figure}%
\centering
\includegraphics[width=0.6\linewidth, angle=-90]{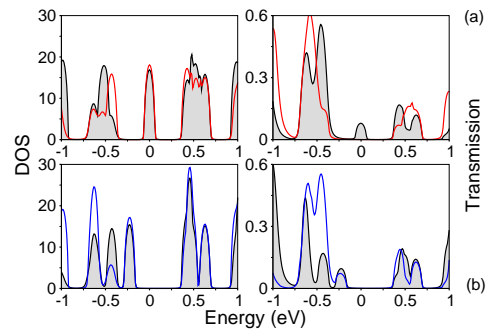}
 \caption{Density of states and Transmission probability of the $n=5$ complex by means of the TB method for nonparameterized (a) and parameterized (b) vacancy values ($Eon-site_{vac}=10eV$, $t_0=1.9 eV$). The DOS figures are represented with a small Gaussian smearing.}
 \label{fig:TBdos}
\end{figure}

Albeit its phenomenological simplicity, the TB description of conduction in the same structures as before generates complexity in the interpretation of the obtained results. The critical points are two: (i) the first has to do with the zero mode introduced by the vacancy at the Fermi energy level. The corresponding HOMO wavefunction, and equally the LDOS at $E=0$, have a succession of finite and zero values for next-neighbor atoms, that is, for each C atom with a finite LDOS value the three nearest neighbor atoms have a zero value and \textit{vice versa}. Moreover the $n=5$ structure (like all even $n$-indexed ones) has a hexagonal side with $LDOS=0$, as discussed in the previous section. Therefore, for the standard-parameterized first-neighbor TB model this state represents the respective molecular orbital in a rigid way, contrary to EH and DFT. (ii) The second issue reflects TB interface bonding issues between a device and the metallic leads that in the next-neighbor context present a strongly-localized character (e.g. only two C atoms in our case are allowed to chemically interact with the leads). In this case, if the metallic contacts form bonding interactions exclusively with zero LDOS carbon atoms (configuration 2 in our case), the HOMO eigenstate will not contribute to the conductivity of the system, yielding a zero transmission probability at that energy. Indeed, figure \ref{fig:TBdos} shows transmission as a function of energy for the $n=5$ complex for the two contact configurations presented before, where a finite transmission probability for the HOMO state appears only in the first case, whereas clearly no conduction takes place through this state for the second. In terms of expectation values, for the HOMO eigenstate of the second configuration we now get:
\begin{equation}
 < \Psi_{HOMO} | \hat{H}_{0R} | \Psi_{HOMO} > = 0
\end{equation} 
This blocked conduction channel reflects the extreme manifestation of wavefunction localization obtained by TB and comes to contrast with EH results. It is therefore fundamental that a realistic modeling has to take into account that the tails of the contact wavefunctions penetrate the body of a molecular device for several \AA{} before they decay.

\begin{figure}%
\centering
\includegraphics[width=0.9\linewidth]{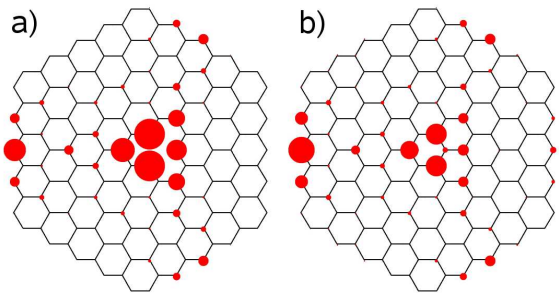}
 \caption{Schematic LDOS representation of the $n=5$ complex by means of the TB model for two different parameterizations of the vacancy site, a) $t_0=0 eV$ within the vacancy and the neighboring sites, and b) $t_0=1.9 eV$ and $Eon-site_{vac}=10eV$. The radius of each circle is proportional to the amplitude of the LDOS value on that atomic site.}
 \label{fig:LDOS}
\end{figure}

A possible way to affront the aforementioned problems is by introducing a further parameterization of the vacancy site. Figure \ref{fig:LDOS} shows a schematic LDOS real-space representation of the $n=5$ island by means of the TB method for a non-parameterized and a parameterized vacancy site. The vacancy parameterization takes place by assigning a finite $E=10eV$ energy on the site and a $t_0=1.9 eV$ hopping integral within this and the neighboring atoms. The principal differentiations obtained are: (i) The sites where zero LDOS values corresponded for the HOMO level now obtain a finite, albeit small density (not visible in fig. \ref{fig:LDOS}). (ii) The hexagonal edge that corresponds to the vacant site (which for even $n$-indexed molecules had zero HOMO-wavefunction components) obtains also a finite LDOS value that is more similar to the electronic structure by means of the DFT and the EH methods (see figure \ref{fig:HOMOvac}). (iii) The collocation of the HOMO eigenstate on the energy axis is in $E<0$, i.e. it moves towards the valance band leaving the midgap position (see figure \ref{fig:TBdos}). Also in this case the DOS spectrum comes closer to the ones obtained by DFT and EH (figure \ref{fig:dos}). Finally, changes obtained for wavefunctions that correspond to other than the HOMO level do not present particular differences from their non parameterized counterparts. Overall, the parameterization of the vacancy site permits for a clear improvement of the qualitative aspects of the electronic structure for defected molecules, respecting the chemical equilibriums and approaching results obtained by more sophisticated methods. Indeed, in terms of transport, the HOMO eigenstate now contributes to conduction for both contact configurations whereas a less `radical' representation of the transmission probability is sketched. Apart from to the qualitative gains of TB calibration presented here though, a key conceptual issue arises. Now the vacancy site becomes similar to a (nominally $p$-type) impurity, since the local point potential lowers (from $\infty$ to big finite) and the hopping integral raises (from $0 eV$ to finite) \cite{pereira:115109}. This consideration can have an impact on the way vacancies are seen in graphene-based systems, both from an applicative as well as from a methodological point of view. As a conclusion, it should be strongly stated that the common perception of treating vacancies in graphene-based systems (zero-energy modes) is not confirmed in this study, whereas an impurity-like behavior has been obtained.

\section{Discussion}
\label{section5}
One important aspect of the understanding of impurity induced disorder in graphene is the possibility of a controllable band gap tailoring for semiconductor applications \cite{2009NanoL...9.2725B,pss}.  This study evidences that vacancies in graphene complexes actually behave as impurities. Such consideration can have a big practical impact on the engineering of mobility gaps in graphene-based systems since vacancies are easier to obtain (e.g. by ion irradiation \cite{compagnini}) than actual $p$ or $n$-type doping. Here we have attempted to affront modeling issues for pure and defected graphene quantum dot islands keeping in mind that the desired computational efficiency for the simulation of large systems should not be in contrast with chemical accuracy. In this sense a multiscale approach has been introduced with the scope to identify merits and limitations of semiempirical approaches within a designated chemical environment prior to their use for the calculation of quantum transport. Model confrontations have demonstrated that no perfect matching exists between the results obtained by the \textit{ab initio} on the one hand and the semiempirical approaches on the other. The extended H\"{u}ckel method with its real-orbital foundation manages to capture a wide set of qualitative aspects of the systems, which qualify it as an appropriate method for quantum transport calculations in graphene-based environments. Moving towards computational efficiency, the tight-binding model has confirmed its authoritativeness for pure large-scale structures, whereas when structural defects have to be accounted a further parameterization of these sites needs to be considered. Unlikely, such tuning cannot be generic for all types of complexes/defect-types since chemical environment influence can be fundamental. E.g. we have to note that by only adding a second vacancy in the vicinity of the same triangular sublattice of the honeycomb structure, hybridization between the two modes can take place. In this sense, a model evaluation of TB by a more sophisticated method should ideally take place prior to its use in disordered graphene-based systems. It should be finally pointed out that both theoretical and experimental attention should be paid to strongly defected systems where topological disorder can be a reason for wavefunction localizations, whose influence on the electronic properties can be important.

\bibliography{revision3}

\end{document}